\documentclass[conference]{IEEEtran}

\IEEEoverridecommandlockouts
\usepackage{graphicx}
\usepackage{cite}
\usepackage{amsmath}
\usepackage{amsfonts}
\usepackage{amssymb}
\usepackage{subfigure}
\usepackage{psfrag}
\usepackage{color}
\usepackage{multicol}

\newcommand{\bm}{\mathbf}

\newcommand{\be}{\begin{equation}}
\newcommand{\ee}{\end{equation}}

\title{Practical Synchronization for OTFS}
\vspace{-0.8cm}
\author{\IEEEauthorblockN{Mohsen Bayat, Sanoopkumar P. S., and Arman Farhang} 
\IEEEauthorblockA{Department of Electronic \& Electrical Engineering, Trinity College Dublin, Ireland \\
\{bayatm, pungayis, arman.farhang\}@tcd.ie}
\vspace{-0.8cm}
\thanks{This publication has emanated from research conducted with the financial support of Science Foundation Ireland under Grant number 19/FFP/7005(T).}}

\begin{document}
\maketitle
\begin{abstract}
In the existing literature on joint timing and frequency synchronization of orthogonal time frequency space modulation (OTFS), practically infeasible impulse pilot with large peak-to-average power ratio (PAPR) is deployed. Hence, in this paper, we propose a timing offset (TO) and carrier frequency offset (CFO) estimation for OTFS over a linear time-varying (LTV) channel, using a low PAPR pilot structure. The proposed technique utilizes the recently proposed practically feasible pilot structure with a cyclic prefix (PCP). We exploit the periodic properties of PCP in both delay and time domains to find the starting point of each OTFS block. Furthermore, we propose a two-stage CFO estimation technique with over an order of magnitude higher estimation accuracy than the existing estimator using the impulse pilot. In the first stage, a coarse CFO estimate is obtained which is refined in the second stage, through our proposed maximum likelihood (ML) based approach. The proposed ML-based approach deploys the generalized complex exponential basis expansion model (GCE-BEM) to capture the time variations of the channel, absorb them into the pilot and provide an accurate CFO estimate. Since our proposed synchronization technique utilizes the same pilot deployed for channel estimation, it does not require any additional overhead. Finally, we evaluate the performance of our proposed synchronization technique through simulations.
We also compare and show the superior performance of our proposed technique to the only other existing joint TO and CFO estimation method in OTFS literature.
\end{abstract}
\vspace{-0.2cm}
\section{Introduction}\label{sec:Introduction}

Orthogonal time-frequency space (OTFS) is a prominent waveform candidate for the sixth-generation (6G) wireless communication systems due to its robustness to time-varying wireless channel and backward compatibility with orthogonal frequency division multiplexing (OFDM), \cite{Hadani2017}. Unlike OFDM, OTFS places modulated data symbols in the delay Doppler (DD) domain and
spreads it on the time-frequency (TF) plane, thus exploiting the full diversity gain of the time and frequency selective channel in high mobility scenarios, \cite{Farhang2018}.  
Since the performance of multicarrier modulations depends on the orthogonality among subcarriers,  timing and carrier synchronization are of paramount importance in modern wireless communication systems.  Even though there exists a large body of work on OFDM synchronization \cite{Morelli2007}, OTFS literature on this topic is still in its infancy.

Timing offset (TO) and/or carrier frequency offset (CFO) estimation in OTFS are addressed in \cite{Sinha2020, Khan2021,Das2021,Bayat2022}.
A threshold-based TO offset estimation technique for OTFS uplink transmission using random access (RA) preamble is presented in \cite{Sinha2020}. In \cite{Khan2021}, the authors developed a correlation-based method for TO estimation in OTFS downlink transmission, which uses a preamble consisting of a linear frequency-modulated (LFM) waveform and two OTFS symbols. Due to the rapid variation of the channel in high mobility scenarios, the TO estimated at the base station using the preamble would be outdated during the data transmission phase. Hence, the methods developed in \cite{Sinha2020} and \cite{Khan2021} are not suitable for high-mobility scenarios, and CFO estimation is not addressed in either \cite{Sinha2020} or \cite{Khan2021}.

A time domain joint channel-CFO estimation and time domain equalization technique for OTFS are presented in \cite{Das2021}. In \cite{Das2021}, the CFO is estimated and compensated as a part of the channel which hinders the pre-compensation of the CFO at the user terminal which is unsuitable for uplink transmissions. Recently, in \cite{Bayat2022}, we addressed the TO and CFO estimation in the OTFS systems, where we developed a correlation-based estimation scheme that exploited the periodic structure of the pilot in the delay-time domain. The embedded impulse pilot which is proposed in \cite{Raviteja2018} and widely used for channel estimation in OTFS systems is employed in \cite{Das2021} and \cite{Bayat2022}.

Accurate TO and CFO synchronization using the methods proposed in \cite{Das2021} and \cite{Bayat2022} requires a high-power impulse pilot. However, the high power of the impulse pilot and the zero guards around it increases the peak-to-average power ratio (PAPR) of the transmitted signal \cite{Sanoop2022}. High PAPR of the transmitted signal will result in a reduction in the efficiency of the power amplifier at the radio frequency (RF) front end, \cite{Gao_2020}. Hence, the widely used impulse pilot is not suitable for practical applications.  To address this issue in \cite{Sanoop2022}, we proposed a novel embedded pilot which uses a constant amplitude pilot sequence with a cyclic prefix (CP), called pilot with cyclic prefix (PCP), placed along multiple delay bins of a single Doppler bin in the delay-Doppler domain. PCP significantly reduces the PAPR of the transmitted signal and thus, it is more suitable for practical applications than impulse pilot.  

Based on the above, the very limited literature available on the synchronization in OTFS uses practically infeasible pilot structures. Hence, to address this issue in this paper, we develop a practical TO and CFO estimation for OTFS, using the PCP deployed for channel estimation in \cite{Sanoop2022}. We exploit the periodicity of the PCP in the delay-time domain and propose a correlation-based TO and coarse CFO estimation. Furthermore, to improve the accuracy of CFO estimation we approximate the time variation of the channel using a generalized complex exponential basis expansion model (GCE-BEM) \cite{Tugnait2010}. We propose a maximum likelihood (ML) CFO fine estimation technique that provides over an order of magnitude higher estimation accuracy than the existing estimator using the impulse pilot. To corroborate our claims, we analyze the performance of our proposed TO and CFO estimation techniques through simulations. In our simulations, we study the mean and variance of the TO estimation error and the mean square error (MSE) of the CFO estimates.

The rest of this paper is organized as follows. Section~\ref{sec:System_Model} describes the system model. The proposed estimation techniques are presented in Sections~\ref{sec:Est} and \ref{sec:CFO_est}, respectively, and their performance is evaluated by simulations in Section~\ref{sec:Simulation_Results}. Finally, Section~\ref{sec:Conclusion} concludes the paper.

$Notations$: Scalar values, vectors, and matrices are denoted by normal letters, boldface lowercase, and boldface uppercase, respectively.
$\rm{diag}[.]$, $\rm{blkdiag}[.]$, $(\!(.)\!)_k$, and $\max_k \{.\}$, present a diagonal matrix, block diagonal matrix, $k^{\rm{th}}$ circular shift, and maximum of a function regarding $k$, respectively.
$|.|$, $\angle .$, and $\rm{Re}\{.\}$ denote the gain, phase, and real part of the complex argument, respectively.
$\bm{I}_{N}$ is an identity matrix with size $N \times N$.
The superscripts $(.)^{\rm{H}}$, $(.)^{\rm{T}}$ and $(.)^{-1}$ indicate hermitian, transpose, and inverse operations, respectively.
$\mathbb{C}^{M \times N}$ stands for a set of complex values with size $M \times N$ and the Kronecker product is denoted by $\otimes$.
Finally, $\mathcal{O}(.)$ presents the order of the complexity for a function. 

\vspace{-0.15cm}
\section{System Model}\label{sec:System_Model}

\begin{figure}[!t]
  \centering 
  {\includegraphics[scale=0.18]{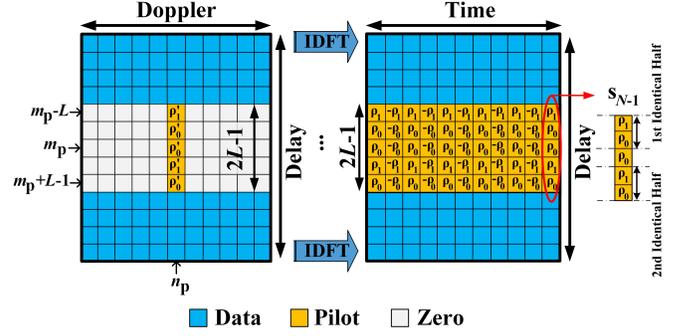}}
  \vspace{-0.9cm}
  \caption{PCP structure in both delay-Doppler and delay-time domains.}
  \vspace{-0.4cm}
  \label{fig1}
  \vspace{-0.2cm}
\end{figure}

We consider an OTFS system with $M$ delay and $N$ Doppler bins, with a Doppler resolution of $\Delta \nu=\frac{1}{MNT_{\rm s}}$ and a delay resolution of $\Delta \tau=T_{\rm s}$, where $T_{\rm s}$ is the sampling period \cite{Hadani2017}. The data symbols and the pilot sequence are multiplexed together to form the delay-Doppler domain OTFS transmitted block $\bm D \in \mathbb{C}^{M\times N}$ with the elements $D[m,n]$ for $m=0,\ldots,M-1$ and $n=0,\ldots,N-1$. 
In this paper, we deploy the pilot structure, PCP, that was originally proposed for channel estimation in \cite{Sanoop2022}, also for synchronization. Considering $L$ as the channel length, in PCP, a constant amplitude Zadoff-Chu (ZC) sequence with length $L$ is placed in the Doppler bin $n_{\rm{p}}$ and the delay bins $m_{\rm{p}},\ldots,m_{\rm{p}}+L-1$. The last $L-1$ samples of this sequence are then appended as a CP in the Doppler bin $n_{\rm{p}}$ and the delay bins $m_{\rm{p}}-L,\ldots,m_{\rm{p}}-1$. The remaining Doppler bins within the pilot region are set to zero.
The delay-Doppler domain OTFS block where the PCP is embedded with data symbols, is shown in Fig.~\ref{fig1}.

The OTFS transmitter spreads the symbols $D[m,n]$ from the delay-Doppler to the delay-time domain by taking inverse discrete Fourier (IDFT) across the Doppler dimension, \cite{Farhang2018},
\be
X[m,l]= \frac{1}{\sqrt{N}} \sum_{n=0}^{N-1} D[m,n] e^{\frac{j2 \pi ln}{N}},
\label{eqn:dt} 
\ee
where $l=0,\ldots, N-1$ is the time index and $m=0\ldots, M-1$ is the delay index.
The delay-time domain signal is then converted to the serial stream $\bm x=[x[0],\ldots,x[MN-1]]^{\rm T}$, where $x[lM+m]=X[m,l]$. Finally, the OTFS transmit signal $s[k]$ is formed by appending a CP with the length $L_{\rm{CP}}\geq L-1$ at the beginning of the OTFS block.
Assuming ideal pulse-shaping, the received delay-time signal for $B$ OTFS blocks, after transmission over the linear time-varying (LTV) channel and in presence of TO and CFO, can be represented as
\be
r[k] = e^{j\frac{2\pi \varepsilon k}{MN}} \sum_{i=0}^{B-1} \sum_{\ell=0}^{L-1} h[\ell,k] s[k-\ell-\theta-iN_{\rm{T}}] + \eta[k], 
\label{eqn:rec} 
\ee
where $0\leq k\leq L_{\rm CP}+MN-1$, $\theta$ and $\varepsilon$ are the TO and CFO values, normalized by the delay and doppler spacings, respectively, and $N_{\rm{T}}=MN+L_{\rm CP}$. $h[\ell,k]$ is the delay-time domain channel impulse response of the $k^{\text{th}}$ delay tap at $\ell^{\text{th}}$ time instant. $\eta[k]$ is the complex additive white Gaussian noise (AWGN) with the variance $\sigma_\eta^2$.

\section{Proposed TO Estimation Technique}\label{sec:Est}

In this section, we present our proposed TO estimation technique for OTFS using PCP. PCP has a very attractive dual periodicity property in the delay-time domain. The periodicity of PCP in the delay dimension is due to the presence of CP. Meanwhile, the periodicity in the time domain is due to the spreading effect of the OTFS transmitter which scales and then repeats each pilot sample across the time dimension.
We exploit this dual periodicity of the PCP in the delay and time dimensions to estimate the TO. 
We consider the TO as $\theta=\theta_{\rm{d}}+M\theta_{\rm{t}}$, where $\theta_{\rm d}$ and $\theta_{\rm t}$  are the TO in delay and time dimensions, respectively. Our proposed TO estimator finds $\theta_{\rm d}$ and $\theta_{\rm t}$ in two stages, without any estimation range limitation.

\vspace{-0.1cm}
\subsection{TO estimation in delay dimension}
The delay-time domain pilot sequence in the delay dimension can be split into two identical halves, each with the length $L-1$, see Fig.~\ref{fig1}.
Assuming that the time variation of the channel within the pilot duration in delay, i.e., $2L-1$ samples, is negligible, the LTV channel over this duration can be considered as linear time-invariant (LTI). Hence, the periodic property of the pilot along the delay dimension is preserved and the TO can be estimated by searching for two identical halves in the received pilot signal. However, channel time variations along the time dimension are not negligible. As it was shown in \cite{Bayat2022}, the identical parts of the pilot along the time dimension should be brought as close as possible to each other to exploit the periodicity in time for TO estimation. The extreme case for this is satisfied for PCP in the delay-time domain as all the pilot samples in a given delay bin $m\in \{m_{\rm p}-L,\ldots,m_{\rm p}+L-1\}$ have the same amplitude and the linear phase of ${{2\pi n_{\rm{p}} l}/{N}}$ for $l=0, 1,\ldots, N-1$. For instance, when $n_{\rm p}=N/2$, the adjacent pilot samples across each delay row have a phase difference of $\pi$, see Fig.~\ref{fig1}. The impulse pilot in \cite{Bayat2022} can only exploit the diversity in $L$ delay bins for TO estimation. In contrast, PCP achieves an improved timing estimation performance, as it takes advantage of the full diversity provided by all the $2L-1$ delay bins allocated to pilot transmission.

\begin{figure}[!t]
  \centering 
  {\includegraphics[scale=0.15]{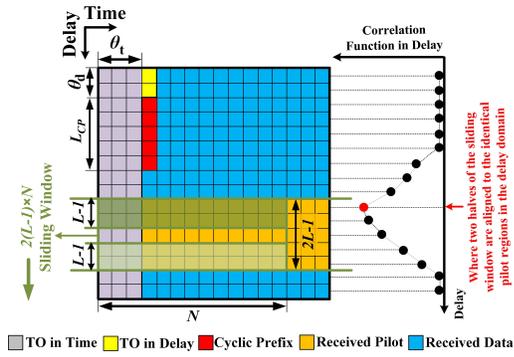}}
  \vspace{-0.4cm}
  \caption{Sliding window for estimation of $\theta_{\rm d}$.}
  \vspace{-0.4cm}
  \label{fig2}
  \vspace{-0.2cm}
\end{figure}

As shown in Fig.~\ref{fig2}, our proposed TO estimator searches for a periodic signal with identical parts in both delay and time dimensions using a sliding window.
To find the periodic sequence in the delay dimension, we define a window with two halves covering $L-1$ samples each. The window slides across the delay dimension to find two pairs of data samples with the highest similarity. In fact, the window searches for the pilot sequences with two identical halves along the delay dimension. To cast this process into a mathematical formulation, using the received signal $r[k]$, we define the timing metric
\be
P_{\rm{d}}[m]= \sum_{i=0}^{N-1} \sum_{u=0}^{L-2} r^*[iM+m+u] r[iM+m+u+L],
\label{eqn:cord}
\ee
that can be efficiently implemented in an iterative manner as
\begin{align}
P_{\rm{d}}[m+1] = P_{\rm{d}}[m] - \sum_{i=0}^{N-1} r^*[iM+m] r[iM+m+1]
\nonumber \\
+ \sum_{i=0}^{N-1} r^*[iM+m+L-2] r[iM+m+L-1],
\label{eqn:itd}
\end{align}
where $m\!=\!0,\ldots,M-1$ and $l\!=\!0,\ldots,N-1$.
Consequently, ${\theta}_{\rm d}$ is estimated by finding the peak of this timing metric as
\be \label{eqn:peakd} 
\hat{\theta}_{\rm d}={\arg} \max_m \{ |P_{\rm{d}}[m]| \}-(m_{\rm{p}}-L)-L_{\rm{CP}}-\lfloor\mu_{{h}}\rfloor,
\ee
where $\mu_{{h}}=\frac{\sum_{\ell=0}^{L-1} (\ell+1) \alpha^2_{\ell}}{\sum_{\ell=0}^{L-1} \alpha^2_{\ell}}$ is the mean of delay that is imposed by the channel and $\alpha_{\ell}$ for $\ell=0,\ldots,L-1$ represents the channel power delay profile (PDP), \cite{Molisch2011}.
The multipath effect of the channel leads to a bias in the TO estimate that can be corrected by the knowledge of the channel's first-order moment \cite{Mensing2007}.
The proposed estimator even works without this knowledge by increasing the CP length with $\lfloor\mu_{{h}}\rfloor$ samples.

\begin{figure}[!t]
  \centering 
  {\includegraphics[scale=0.15]{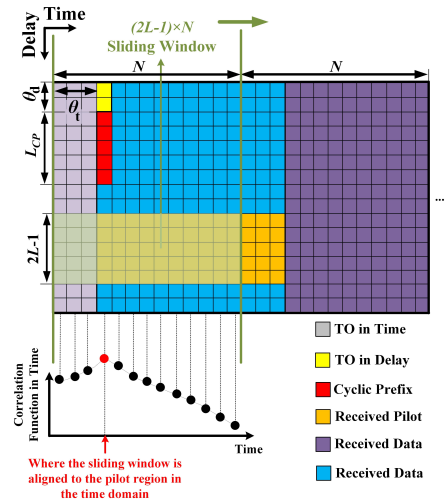}}
  \vspace{-0.4cm}
  \caption{Sliding window for estimation of $\theta_{\rm t}$.}
  \vspace{-0.4cm}
  \label{fig3}
  \vspace{-0.2cm}
\end{figure}

\subsection{TO estimation in time dimension}
The peak of the correlation function $P_{\rm{d}}[m]$ on the row $m'_{\rm p}-L=\hat{\theta}_{\rm d}+(m_{\rm p}-L)+L_{\rm{CP}}+\lfloor \mu_{h} \rfloor$ of the delay-time grid can provide an estimate of $\theta_{\rm t}$, where $m'_{\rm p}=\hat{\theta}_{\rm d}+m_{\rm p}$. However, this estimate is not very accurate. This is while even a single error in time cannot be afforded as the estimation error of one sample in $\theta_{\rm t}$ leads to an effective  error of $M$ samples in the final TO estimate. This highlights the importance of the need for a highly accurate estimation of $\theta_{\rm t}$. 
Thus, to estimate $\theta_{\rm t}$, we deploy a sliding window with the length $2N-1$ that covers the delay bins $m'_{\rm{p}}-L,\ldots,m'_{\rm{p}}+L-1$ and slides along time, see Fig.~\ref{fig3}. This window calculates the correlation between every two adjacent samples in time for all $2L-1$ delay bins in the pilot region.
This process can be mathematically shown as
\be
P_{\rm{t}}[l]= \sum_{i=m'_{\rm{p}}-L}^{m'_{\rm{p}}+L-1} \sum_{v=0}^{N-2} r^*[(l+v)M+i] r[(l+v+1)M+i],
\label{eqn:cort}
\ee
that can be iteratively implemented as
\begin{align}
P_{\rm{t}}[l+1] = P_{\rm{t}}[l] - \sum_{i=m'_{\rm{p}}-L}^{m'_{\rm{p}}+L-1} r^*[lM+i] r[(l+1)M+i]
\nonumber \\
+ \sum_{i=m'_{\rm{p}}-L}^{m'_{\rm{p}}+L-1} r^*[(l+N-2)M+i] r[(l+N-1)M+i],
\label{eqn:itt}
\end{align}
where $l=0,1,...,N-1$.
Hence, $\theta_{\rm{t}}$ is estimated by finding the peak of $P_{\rm{t}}$, i.e., 
\be
\hat{\theta}_{\rm t}={\arg} \max_l \{ |P_{\rm{t}}[l]| \}.
\label{eqn:peakt}
\ee

\begin{figure}[!t]
  \centering 
  {\includegraphics[scale=0.35]{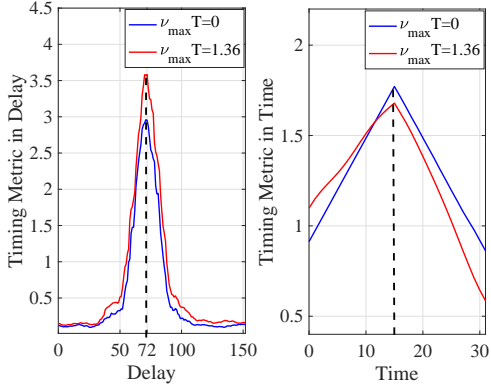}}
  \vspace{-0.3cm}
  \caption{One snapshot of the timing metrics in delay and time dimensions for SNR=20~dB where $(\theta_{\rm{d}}+L_{\rm{CP}})=72$ and $\theta_{\rm{t}}=15$.}
  \vspace{-0.4cm}
  \label{fig4}
  \vspace{-0.2cm}
\end{figure}

Fig.~\ref{fig4} shows a snapshot of the timing metrics at the SNR of $20$~dB for an OTFS system with $M=128$ and $N=32$ for both LTI and LTV channels. 

After correcting the TO with $\hat{\theta}=\hat{\theta}_{\rm d}+M \hat{\theta}_{\rm t}$, in the following section, we propose a novel two-stage CFO estimation technique. Our proposed technique finds a coarse estimate of the CFO by using the angle of the timing metric that we used for TO estimation, and then this estimate is refined by using our proposed ML estimation technique.

\section{CFO Estimation Using Maximum Likelihood}\label{sec:CFO_est}
\begin{figure*}[h]
 \be
 \tag{13}
 \begin{aligned}
 \bm{H}_l =\!\!
 \begin{bmatrix}
 h[0,L_{\rm{CP}}+lM+m_{\rm{p}}] &  h[L\!-\!1,L_{\rm{CP}}+lM+m_{\rm{p}}]  & \cdots & h[1,L_{\rm{CP}}+lM+m_{\rm{p}}]
\\
h[1,L_{\rm{CP}}+lM+m_{\rm{p}}+1] & h[0,L_{\rm{CP}}+lM+m_{\rm{p}}+1] & \cdots & h[2,L_{\rm{CP}}+lM+m_{\rm{p}}+1]
\\
\vdots &\vdots &\ddots &\vdots
\\
h[L-1,L_{\rm{CP}}+lM+m_{\rm{p}}+L\!-\!1] & h[L-2,L_{\rm{CP}}+lM+m_{\rm{p}}+L\!-\!1] & \cdots & h[0,L_{\rm{CP}}+lM+m_{\rm{p}}+L\!-\!1]
\end{bmatrix}
\label{eqn:ch2}
\end{aligned}
\ee
\vspace{-0.95cm}
\end{figure*}
Considering this phase difference, and averaging the correlation angle over the delay bins allocated with the pilot at the timing instant $\hat{\theta}_{\rm t}$ from (\ref{eqn:cort}), 
\begin{align}
\Upsilon_{\hat{\theta}_{\rm t}}= \!\frac{1}{2L-1} \sum_{i=m'_{\rm{p}}-L}^{m'_{\rm{p}}+L-1}  {\angle} \!\sum_{v=0}^{N-2} &( r^*[(\hat{\theta}_{\rm t}+v)M+i] \times \nonumber \\ 
&r[(\hat{\theta}_{\rm t}+v+1)M+i]),
\end{align}
a coarse CFO estimate can be obtained as
 \be \label{eqn:cfo1} 
\hat{\varepsilon}_{\rm c}=\frac{N}{2\pi} \Upsilon_{\hat{\theta}_{\rm t}}-n_{\rm{p}}.
 \ee
To refine this CFO estimate and improve the estimation accuracy, in the following, we develop an ML-based technique as a fine CFO estimation stage.  

After correcting the TO, the received pilots in the delay bins $m_{\rm{p}}$ to $m_{\rm{p}}+L-1$ over all the bins along the time dimension are used for CFO estimation. For ease of explanation, in the rest of the paper, the received pilots refer to the received signal in the delay bins $m_{\rm{p}}$ to $m_{\rm{p}}+L-1$. After removing the CP from the received pilot, and stacking the resulting signals at different time slots, $\bm{r}_{l,\rm p}=[r[L_{\rm CP}+lM+m_{\rm p}],\ldots,r[L_{\rm CP}+lM+m_{\rm p}+L-1]]^{\rm T}$, into the vector $\bm{r}^{\rm{p}}=[\bm{r}_{0,\rm p}^{\rm T},\bm{r}_1^{\rm T},\ldots,\bm{r}_{N-1,\rm p}^{\rm T}]^{\rm{T}}\in \mathbb{C}^{NL\times 1}$, using (\ref{eqn:rec}), $\bm{r}^{\rm{p}}$ can be expressed as, 
\be
\bm{r}^{\rm{p}} = \bm{\Gamma}(\varepsilon) \bm{H} \bm{s}^{\rm{p}} + \boldsymbol{\eta}, 
\label{eqn:recM} 
\ee
where, $\bm{\Gamma(\varepsilon)}\!=\!{\rm{blkdiag}} [\bm{\Gamma}_0,\bm{\Gamma}_1,\ldots,\bm{\Gamma}_{N-1}]$ with 
\begin{align}
\bm{\Gamma}_l={\rm{diag}}[e^{\frac{j2\pi \varepsilon (L_{\rm{CP}}+lM+m_{\rm{p}})}{N_{\rm{T}}}},e^{\frac{j2\pi \varepsilon (L_{\rm{CP}}+lM+m_{\rm{p}}+1)}{N_{\rm{T}}}},\ldots, \nonumber\\
e^{\frac{j2\pi \varepsilon (L_{\rm{CP}}+lM+m_{\rm{p}}+L-1)}{N_{\rm{T}}}}],
\label{eqn:cfo}
\end{align}
and $\bm{H}\!=\!{\rm{blkdiag}}[\bm{H}_0,\bm{H}_1,...,\bm{H}_{N-1}]$ with $\bm{H}_l$ being the channel matrix, whose structure is shown in (\ref{eqn:ch2}), on the top of the next page.
In (\ref{eqn:recM}), $\bm{s}^{\rm{p}}=[\bm{s}_{0,{\rm p}}^{\rm T},\ldots,\bm{s}_{N-1,{\rm p}}^{\rm T}]^{\rm{T}}\in \mathbb{C}^{NL\times 1}$, $\bm{s}_{l,{\rm p}}$ is the transmitted delay-time pilot samples in the delay bins $m_{\rm{p}}$ to $m_{\rm{p}}+L-1$ and the time slot $l$. $\boldsymbol{\eta} \sim \mathcal{CN} (\bm{0}, \bm{\sigma}_\eta^2 \bm{I}_{LN})$ is the AWGN vector with size $LN \times 1$ that affects the pilot.
By interchanging the convolution order in (\ref{eqn:recM}), the received pilot at the time slot $l$ can be expressed as
\be 
\tag{14}
\bm{r}_{l,{\rm p}} = \bm{\Gamma}_l(\varepsilon) \bm{A}_{l,{\rm p}} \bm{h}_l + \boldsymbol{\eta}_l, 
\label{eqn:recM2}
\ee
where $\bm{A}_{l,{\rm p}}=[\bm{S}_{l,{\rm p}}^0,...,\bm{S}_{l,{\rm p}}^{L-1}]$, $\bm{S}_{l,{\rm p}}^{\ell}={\rm{diag}}[(\!(\bm{s}_{l,{\rm p}})\!)_{\ell}]$, $\bm{h}_l=[(\bm{h}_l^{0})^{\rm{T}},(\bm{h}_l^{1})^{\rm{T}},\ldots,(\bm{h}_l^{L-1})^{\rm{T}}]^{\rm{T}}$, and $\bm{h}_l^{\ell}=[h[\ell, L_{\rm{CP}}+lM+m_{\rm{p}}],h[\ell, L_{\rm{CP}}+lM+m_{\rm{p}}+1],\ldots,h[\ell, L_{\rm{CP}}+lM+m_{\rm{p}}+L-1]]^{\rm{T}}$ for $\ell=0,1,\ldots,L-1$. $\boldsymbol{\eta}_l \in \mathbb{C}^{L\times 1}$ is the AWGN at time slot $l$ affecting the pilot. 

BEM-based methods are used for approximating the time variation of the LTV channels. Depending on the basis functions, different BEM methods  such as complex exponential based (CE-BEM)\cite{Fengzhong_2008}, polynomial based BEM \cite{Muneer_2015}, Karhunen-Loeve (KL) decomposition based BEM \cite{Zhang_2006}, are presented in the literature. Due to its simplicity and high accuracy, in this paper, we use the oversampled GCE-BEM \cite{Tugnait2010} to approximate the channel time variation in the delay-time domain. The channel coefficient for the $\ell^{\text{th}}$ path at the time instant $k$ can be expressed using GCE-BEM as
\be
\tag{15}
h[\ell,k]=\sum_{q=0}^{Q-1}B[k,q]c_{\ell}(q),
\label{eqn:bem1}
\ee
where $B[k,q]\!=\!e^{j2\frac{2\pi(q-\lceil Q/2 \rceil)k}{KMN}}$, $0 \!\leq\! k \!\leq\! MN\!-\!1$, $0 \!\leq\! \ell \!\leq\! L\!-\!1 $ and $1\leq q \leq Q$. For the accurate approximation of the time variation of the channel, the oversampling factor and the number of basis functions are chosen as $K\!\geq \!1$ and $Q=\lceil 2K \nu_{\rm{max}}(MNT_{\rm s})\rceil+1$, respectively, \cite{Tugnait2010}.
Using (\ref{eqn:bem1}), $\bm{h}_l$ can be expressed in terms of BEM coefficients as 
\be
\tag{16}
\bm{h}_l=(\bm{I}_L \otimes\bm{B}^{\rm p}_l) \bm{c},
\label{eqn:bem2}
\ee
where $\bm{B}^{\rm p}_l=B[k,q]\, \forall k\in\lbrace L_{\rm{CP}}+lM+ m_{\rm p}, L_{\rm{CP}}+lM+m_{\rm p}+1, \hdots, L_{\rm{CP}}+lM+m_{\rm p}+L-1\rbrace$, $\bm{c}=[\bm{c}_0^{\rm T},\bm{c}_1^{\rm T}, \hdots, \bm{c}_{L-1}^{\rm T}]^{\rm T}$ and $\bm{c}_{\ell}=[c_{\ell}(0),c_{\ell}(1), \hdots,c_{\ell}(Q-1)]^{\rm T}$. Inserting, (\ref{eqn:bem2}) in (\ref{eqn:recM2}), $\bm{r}^{\rm p}$ in (\ref{eqn:recM}) can be approximated using GCE-BEM as
\be
\setcounter{equation}{17}
{\bm{r}}^{\rm{p}} = {\bm{\Gamma}}(\varepsilon) {\bm{G}} {\bm{c}} + {\boldsymbol{\eta}}, 
\label{eqn:recM3} 
\ee
where $\bm{G}=[{\bm G}_0^{\rm T}, {\bm G}_1^{\rm T}, \ldots, {\bm G}_{N-1}^{\rm T}]^{\rm T}$ and ${\bm G}_l={\bm A}_l (\bm{I}_L \otimes\bm{B}^{\rm p}_l)$ for $l=0,1,\ldots,N-1$.

For a given pair $(\bm{c},\varepsilon)$, the vector $\bm{r}^{\rm{p}}$ is assumed to have the Gaussian distribution with the mean 
$\bm{\Gamma}(\varepsilon) \bm{G} \bm{c}$ and covariance matrix $\bm{\sigma}_\eta^2 \bm{I}_{LN}$, \cite{Morelli2000}. Thus, the joint probability density function of $\bm{r}^{\rm{p}}$, parameterized by $(\tilde{\bm{c}},\tilde{\varepsilon})$, is given by
\be
f(\bm{r}^{\rm{p}};\tilde{\bm{c}},\tilde{\varepsilon}) = \frac{1}{(\pi \sigma_{\rm{\eta}}^2)^{NL}} e^{\frac{-1}{\sigma_{\rm{\eta}}^2} [\bm{r}^{\rm{p}}-\bm{\Gamma}(\tilde{\varepsilon}) \bm{G} \tilde{\bm{c}}]^{\rm{H}} [\bm{r}^{\rm{p}}-\bm{\Gamma}(\tilde{\varepsilon}) \bm{G} \tilde{\bm{c}}]}.
 \label{eqn:ML} 
 \ee
Thus, the ML estimates of the BEM coefficient vector and the CFO are obtained as 
\be
(\hat{\bm{c}},\hat{\varepsilon})=\arg \max_{\tilde{\bm{c}},\tilde{\varepsilon}} \{ f({\bm r}^{{\rm p}};\tilde{\bm{c}},\tilde{\varepsilon}) \}.
\label{eqn:ml1}
\ee

Taking the logarithm and removing the constant terms, the estimation problem in (\ref{eqn:ml1}) can be simplified as 
\be
(\hat{\bm{c}},\hat{\varepsilon})=\arg \max_{\tilde{\bm{c}},\tilde{\varepsilon}} \{ g(\tilde{\bm{c}},\tilde{\varepsilon}) \},
\label{eqn:ml2}
\ee
where $g(\tilde{\bm{c}},\tilde{\varepsilon})=\frac{-1}{\sigma_{\rm{\eta}}^2} [\bm{r}^{\rm{p}}-\bm{\Gamma}(\tilde{\varepsilon}) \bm{G} \tilde{\bm{c}}]^{\rm{H}} [\bm{r}^{\rm{p}}-\bm{\Gamma}(\tilde{\varepsilon}) \bm{G} \tilde{\bm{c}}]$ is the joint cost function. The maximization problem in (\ref{eqn:ml2}) can be solved in two steps. In step 1, we find the $\tilde{\bm{c}}$ which maximizes the joint cost function parameterized by $\tilde{\varepsilon}$. In step 2, the $\tilde{\bm{c}}$ obtained in step 1 is used to find a new cost function for $\tilde{\varepsilon}$ and we perform a grid search in the vicinity of coarse CFO estimate to find the fine CFO estimate which maximizes the cost function for $\tilde{\varepsilon}$.
Thus, we fix the $\tilde{\varepsilon}$ and the $\tilde{\bm{c}}$ that maximizes $g(\tilde{\bm{c}},\tilde{\varepsilon})$ can be obtained as 
\be
\tilde{\bm{c}}(\tilde{\varepsilon}) = (\bm{G}^{\rm{H}} \bm{G})^{-1} \bm{G}^{\rm{H}} \bm{\Gamma}^{\rm{H}}(\tilde{\varepsilon}) \bm{r}^{\rm{p}}.
\label{eqn:ch} 
\ee
Substituting $\tilde{\bm{c}}(\tilde{\varepsilon})$ into $g(\tilde{\bm{c}},\tilde{\varepsilon})$, the cost function for CFO can be obtained as 
\be
g_{\rm {CFO}}(\tilde{\varepsilon}) = (\bm{r}^{\rm{p}})^{\rm{H}} \bm{\Gamma}(\tilde{\varepsilon}) \bm{\Lambda} \bm{\Gamma}^{\rm{H}}(\tilde{\varepsilon}) \bm{r}^{\rm{p}},
\label{eqn:cost} 
\ee
where $\bm{\Lambda}=\bm{G} (\bm{G}^{\rm{H}} \bm{G})^{-1} \bm{G}^{\rm{H}}$.
The fine estimate of CFO can then be obtained using a single-dimensional search centered around the coarse CFO estimate $\hat{\varepsilon}_{\rm c}$ in  (\ref{eqn:cfo1}) as
\be
\hat{\varepsilon}={\arg} \max_{\tilde{\varepsilon}} \{g_{\rm{CFO}}(\tilde{\varepsilon})\}. 
\label{eqn:cost2} 
\vspace{-0.1cm}
\ee
In addition, channel estimation can also be developed using the estimated CFO. The BEM coefficients can be estimated after obtaining the CFO estimate as, $\hat{\bm{c}} = (\bm{G}^{\rm{H}} \bm{G})^{-1} \bm{G}^{\rm{H}} \bm{\Gamma}^{\rm{H}}(\hat{\varepsilon}) \bm{r}^{\rm{p}}$, and finally the complete LTV channel in delay-time can be estimated using (\ref{eqn:bem1}).

\begin{figure}[!t]
  \centering 
  {\includegraphics[scale=0.35]{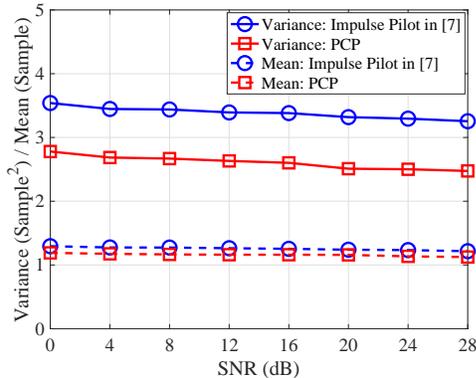}}
  \vspace{-0.4cm}
  \caption{TO estimation comparison for the impulse pilot and PCP where $M\!\!=\!\!128$ and $N\!\!=\!\!32$.}
  \vspace{-0.4cm}
\label{fig5}
\end{figure}

Regarding complexity, the periodic structure of the pilot in the delay-time domain can reduce the complexity of the maximum-likelihood estimator. In other words, the existing repetition of every $L$ pilot sample leads to a reduction of the complexity of the estimator by a factor of $L$.
Additionally, $\bm{\Lambda}$ is a symmetric matrix that provides the opportunity to only calculate half of the cost function, which reduces the complexity by a factor of 2. Thus, the cost function in (\ref{eqn:cost}) can be calculated in the form
\vspace{-0.2cm}
\be
g_{\rm {CFO}}(\tilde{\varepsilon}) = -\beta[0]+2{\rm{Re}} \{ \sum_{m=0}^{N-1} \beta[m] e^{\frac{j2 \pi m \tilde{\varepsilon}}{N}} \},
\label{eqn:cost3}
\vspace{-0.2cm}
\ee
and
\be
\beta[m] =\!\! \sum_{k=0}^{NL-1-mL} \!\! \bm{\Lambda}[(\!(k+mL)\!)_{NL},k] {r^{\rm{p}}}^*[(\!(k+mL)\!)_{NL}]
r^{\rm{p}}[k] ,
\label{eqn:costs}
\ee
where $\bm{\Lambda}[i,j]$ and $r^{\rm{p}}[i]$ are the $[i,j]^{\rm{th}}$ and $i^{\rm{th}}$ entries of the $\bm{\Lambda}$ and $\bm{r}^{\rm{p}}$, respectively. Calculating the coast function $g_{\rm {CFO}}(\tilde{\varepsilon})$ using (\ref{eqn:cost}) requires $\mathcal{O}(N^2L^2)$ complex multiplications. However, it can be reduced to $\mathcal{O}(\frac{N^2L}{2})$ by using (\ref{eqn:cost3}) and (\ref{eqn:costs}).

\vspace{-0.5cm}
\section{Simulation Results}\label{sec:Simulation_Results}
\begin{figure}[!t]
  \centering 
  {\includegraphics[scale=0.35]{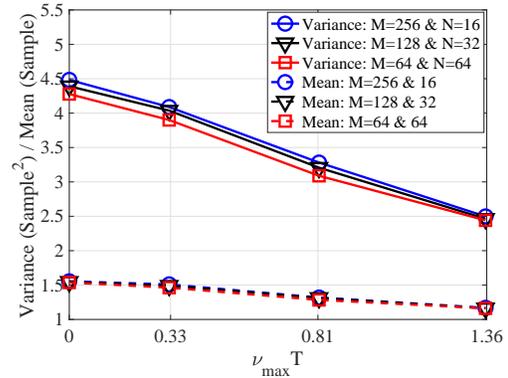}}
  \vspace{-0.4cm}
  \caption{TO estimation comparison vs. normalized maximum Doppler spread for PCP at SNR=20~dB.}
  \vspace{-0.4cm}
\label{fig6}
  \vspace{-0.2cm}
\end{figure}
In this section, we numerically analyze the performance of both the proposed estimation techniques. The reduced-OTFS system with 16-QAM symbols in this analysis is formed by $M=64, 128, 256$ and $N=64, 32, 16$ delay and Doppler bins, respectively \cite{Raviteja2019b}. 
We use the extended vehicular A (EVA) channel model with length $L=21$, \cite{3gpp}, the bandwidth of $8.25$~MHz and the delay-Doppler resolution $(\Delta \tau,\Delta \nu)=(121.21 \, \rm{nsec},2.01 \, \rm{kHz})$.
The power of the PCP is set to 40~dB and the pilot is inserted at the center of the Doppler axis within delay bins, $m_{\rm{p}}-L$ to $m_{\rm{p}}+L-1$, \cite{Sanoop2022}.
To model the channel with GCE-BEM, we choose the oversampling factor $K=4$ and the order of BEM basis functions $Q=1,3,6,8$ for different normalized maximum Doppler spreads of $\nu_{\rm{max}}=0, 0.66, 1.64, 2.73$ kHz, respectively \cite{Tugnait2010}.
Throughout our simulations, the normalized TO and CFO values are randomly generated from a uniform distribution in the range $[-\frac{MN}{2},\frac{MN}{2})$ and $[-\frac{N-\nu_{\rm max}T}{2},\frac{N-\nu_{\rm max}T}{2})$, respectively, where $T=MNT_{\rm{s}}$ is the total time duration of an OTFS block.

In Fig.~\ref{fig5}, we compare the performance of our proposed TO estimation technique using PCP with the impulse pilot-assisted method proposed  in \cite{Bayat2022}. We analyze the estimation error mean and variance as a function of signal-to-noise ratio (SNR), for the normalized Doppler spread of $\nu_{\rm{max}}T\!\approx\! 1.36$.
The PDP of the channel leads to a small bias in the proposed estimation technique. Although we removed $\lfloor \mu_{\rm{h}} \rfloor$ from the estimated TO in (\ref{eqn:peakd}), the existing bias in Fig.~\ref{fig5} originates from the fractional part of the mean delay in the channel.
This figure also shows that, for the same pilot power level, the proposed technique outperforms the technique in  \cite{Bayat2022}.
Fig.~\ref{fig6} shows the performance of the proposed TO estimator as a function of normalized Doppler spread, for different combinations of $M$ and $N$, and a fixed $MN=4096$. It can be observed that  as the normalized Doppler spread increases,  estimation accuracy also increases. This improvement originates from the diversity provided by the time-selectivity of the channel.

\begin{figure}[!t]
  \centering 
  {\includegraphics[scale=0.35]{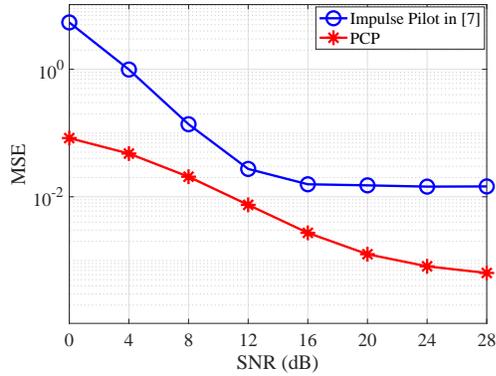}}
  \vspace{-0.4cm}
  \caption{CFO estimation comparison for the impulse pilot and PCP where $M\!\!=\!\!128$ and $N\!\!=\!\!32$.}
  \vspace{-0.4cm}
  \label{fig7}
  \vspace{-0.2cm}
\end{figure}

\begin{figure}[!t]
  \centering 
  {\includegraphics[scale=0.35]{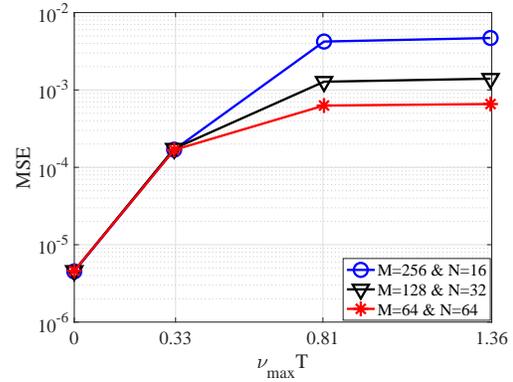}}
  \vspace{-0.4cm}
  \caption{CFO estimation comparison vs. normalized maximum Doppler spread for PCP at SNR=20~dB.}
  \vspace{-0.6cm}
  \label{fig8}
\end{figure}

To analyze the performance of our proposed CFO estimation technique, we assume perfect knowledge of TO. In Fig.~\ref{fig7}, we evaluate the MSE performance of our proposed CFO estimation technique as a function of SNR and compare it with the CFO estimation technique proposed  in \cite{Bayat2022}. It can be observed that the  CFO estimation proposed in this paper gives an order of magnitude better estimation accuracy than the method using impulse pilot in \cite{Bayat2022}. 
In Fig.~\ref{fig8}, we study the performance of our proposed CFO estimator as a function of Doppler spread. 
Furthermore, our results show that the MSE performance degrades as the Doppler spread or the number of delay bins increases. 
Higher $M$ leads to a channel with more time variation  within a column of the OTFS block which is the main reason for CFO estimation degradation.

\section{Conclusion}\label{sec:Conclusion}

In this paper, we proposed TO and CFO estimation techniques for OTFS using PCP. We exploit the periodicity property of PCP in the time domain to develop the correlation-based metrics for TO and CFO estimation. Furthermore, the TO estimation accuracy is improved using diversity offered by the constant amplitude PCP in different delay bins. To improve the accuracy of the CFO estimate we approximated the time variation of the channel using GCE-BEM and developed an ML estimator. Additionally, we also developed low-complexity implementation strategies for the proposed techniques. Since we use the same practical pilot used for channel estimation, the proposed method does not impart additional overhead for synchronization. Hence the proposed synchronization methods are highly apt for practical OTFS systems in high mobility scenarios in the envisioned 6G systems. 

\bibliographystyle{IEEEtran} 
\bibliography{IEEEabrv,references}

\end{document}